\documentclass[final, conference, a4paper]{IEEEtran}
\IEEEoverridecommandlockouts
\usepackage{cite}
\usepackage{amsmath,amssymb,amsfonts}
\usepackage{algorithmic}
\usepackage{graphicx}
\usepackage{textcomp}
\usepackage{hyperref}
\def\BibTeX{{\rm B\kern-.05em{\sc i\kern-.025em b}\kern-.08em
    T\kern-.1667em\lower.7ex\hbox{E}\kern-.125emX}}
\begin{document}
	
\begin{titlepage}
	\begin{center}
		
		\Huge
		\textbf{Detection of direct path component absence \\ in NLOS UWB channel
		}
		
		\vspace{0.5cm}
		\LARGE
		Accepted version
		
		\vspace{1.5cm}
		
		\text{Marcin Kolakowski, Jozef Modelski}
		
		\vspace{.5cm}
		\Large
		Institute of Radioelectronics and Multimedia Technology
		
		Warsaw University of Technology
		
		Warsaw, Poland,
		
		contact: marcin.kolakowski@pw.edu.pl

		\vspace{2cm}

	\end{center}
	
	\Large
	\noindent
	\textbf{Originally presented at:}
	
	\noindent
	2018 22nd International Microwave and Radar Conference (MIKON), Poznan, Poland, 2018
	
	\vspace{.5cm}
	\noindent
	\textbf{When using the paper, please cite as:}
	
	\noindent
	M. Kolakowski and J. Modelski, "Detection of direct path component absence in NLOS UWB channel," 2018 22nd International Microwave and Radar Conference (MIKON), Poznan, Poland, 2018, pp. 247-250, doi: 10.23919/MIKON.2018.8405190
	
	\vspace{.5cm}
	\noindent
	\textbf{Full version available at:}
	
	\noindent
	\url{https://doi.org/10.23919/MIKON.2018.8405190}

	\vspace{.5cm}
	\noindent
	\textbf{Additional information:}
	
	\noindent
	The dataset used in the study is available at Zenodo:
	
	\noindent
Marcin Kolakowski. (2021). UWB Channel Impulse Responses Registered in a Furnished Apartment (Version 1.0) [Data set]. Zenodo. \url{http://doi.org/10.5281/zenodo.4742391}
	
	\vfill
	
	\large
	\noindent
	© 2018 IEEE. Personal use of this material is permitted. Permission from IEEE must be obtained for all other uses, in any current or future media, including reprinting/republishing this material for advertising or promotional purposes, creating new collective works, for resale or redistribution to servers or lists, or reuse of any copyrighted component of this work in other works.
\end{titlepage}

\bibliographystyle{IEEEtran}
\title{Detection of Direct Path Component Absence in~NLOS UWB Channel\\
\thanks{This work was partly supported by the National Centre for Research and Development, Poland under Grant AAL/Call2016/3/2017 (IONISproject).}
}

\author{\IEEEauthorblockN{Marcin Kolakowski}
\IEEEauthorblockA{\textit{Institute of Radioelectronics and Multimedia Technology} \\
\textit{Warsaw University of Technology}\\
Warsaw, Poland \\
m.kolakowski@ire.pw.edu.pl}
\and
\IEEEauthorblockN{Jozef Modelski}
\IEEEauthorblockA{\textit{Institute of Radioelectronics and Multimedia Technology} \\
\textit{Warsaw University of Technology}\\
Warsaw, Poland \\
j.modelski@ire.pw.edu.pl}
}

\maketitle

\begin{abstract}
In this paper a novel NLOS (Non-Line-of-Sight) identification technique is proposed. In comparison to other methods described in the literature, it discerns a situation when the delayed direct path component is available from when it's totally blocked and introduced biases are much higher and harder to mitigate.

In the method, NLOS identification is performed using Support Vector Machine (SVM) algorithm based on various signal features. The paper includes description of the method and the results of performed experiment.
\end{abstract}


\section{Introduction}
Ultra-wideband positioning systems are among most accurate solutions for indoor localization. Their ability to conduct precise time measurements allows to localize objects with low uncertainty down to several dozen centimeters. This makes them a perfect choice for many industrial applications  requiring precise localization including personnel tracking or  production automation.  Unfortunately such high accuracy is  attainable only under LOS  (Line-of-Sight) operating conditions, which are very hard to provide in most typical densely cluttered environments. Therefore most of the time the systems are forced to work in NLOS (Non-Line-of-Sight) conditions.

Signals propagating in NLOS conditions are usually delayed and attenuated, which makes time of arrival measurements performed in the systems biased and less precise due to their higher variance. In order to minimize those effects various NLOS identification and mitigation methods have been developed and presented in the literature.
The proposed methods range from simple methods working in real-time and requiring scarce information about system surroundings to advanced methods for which earlier calibration is needed.

The simplest identification methods rely mostly on power measurements which are made by UWB radio modules. Thanks to their usually low computational complexity they can be succesfully used in real time.  In \cite{gururaj_real-time_2017-1}  NLOS conditions are identified by comparing a difference between total and first path signal power with a predefined threshold. The method proposed in \cite{albaidhani_ranging_2016-1} defines two thresholds and additionally analyzes total power, which makes the method more robust.

More advanced methods consist in registering received signal waveforms or channel impulse responses and extracting from them features, which usually include kurtosis, mean excess delay, root mean square delay spread, signal amplitude and other various parameters. The collected features are then the basis for NLOS conditions identification.

Many methods of NLOS identifications involve hypothesis testing \cite{silva_ir-uwb-based_2016-1} but recently machine learning classification techniques  gain more recognition. The most popular algorithms used for that purpose are Support Vector Machine (SVM) \cite{marano_nlos_2010-1},\cite{miao_application_2016-1},\cite{li_experimental_2013-1} and k-Nearest Neighbors (kNN) \cite{zhang_low_2015-1}. Those methods allow to identify NLOS conditions with very high accuracy but are computationally demanding  and require earlier data collection for training, which may pose a problem in some applications.

A very important issue in NLOS propagation is the availability of direct path component. In case when it is available, introduced biases are much lower and easier to mitigate than in case when the systems works only with the reflected components. Although some researches \cite{gururaj_real-time_2017-1} and \cite{silva_ir-uwb-based_2016-1} recognize the difference between these both situations, they do not propose methods for the lack of direct path component identification. In \cite{gururaj_real-time_2017-1} the authors focus on identifying propagation through obstacles from LOS conditions, whereas in \cite{silva_ir-uwb-based_2016-1} the proposed method does not differentiate between different NLOS scenarios. 

In the paper a novel method for NLOS identification is proposed. In comparison to previously mentioned techniques it recognizes whether a direct path component is available or not. The proposed method identifies NLOS conditions based on  signal features extracted from recorded waveforms and power measurements using Support Vector Machine (SVM) algorithm. The remainder of the paper is organized as follows. A more specific description of the problem is provided in section \ref{formulation}. In sections \ref{approach} and \ref{ML} NLOS identification method is proposed. Experiment results are presented in section \ref{exp}. Section \ref{con} concludes the paper.

\section{Problem formulation} \label{formulation}
\begin{figure}[!b]
\centerline{\includegraphics[width=3.35 in]{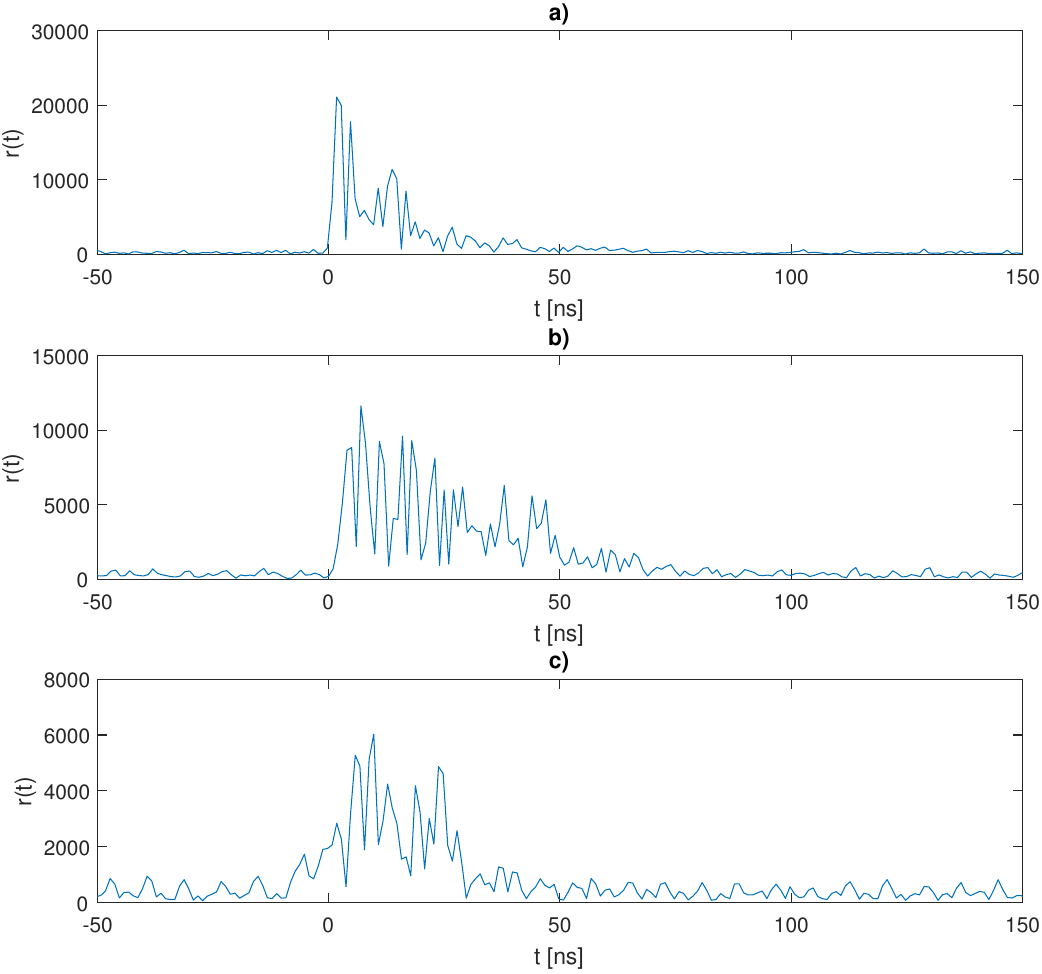}}
\caption{Signal magnitude recorded by DW1000 radio chip \cite{decawave_ltd_dw1000_2015} in: a) LOS conditions, b)~NLOS conditions with received Direct Patch Component, c) NLOS conditions in absence of Direct Path Component. Time 0 ns corresponds to the moment of first component detection.}
\label{fig:CIRs}
\end{figure}

Signals propagating in NLOS conditions are usually delayed and  attenuated. It results in that obtained time of arrival or ranging measurement results are biased and have higher variance. The degree to which the results are deteriorated depends on  the type of obstacles which are blocking the direct visibility between the system devices. A comparison of signals obtained under LOS and two different types of NLOS conditions is presented in Fig.\ref{fig:CIRs}.

When the direct path component propagating through the obstacle is available as in Fig.\ref{fig:CIRs}b, the introduced bias usually would not exceed 2 ns. From values given in \cite{albaidhani_ranging_2016-1}, it can be estimated that propagation through two 10 cm thick concrete walls would result in about 1 ns delay. It has been shown that such moderate bias values can be successfully mitigated \cite{albaidhani_ranging_2016-1}. 

In case of harsher NLOS propagation conditions, in which the direct path component is not available or is its level is too low to be properly received and detected (Fig.\ref{fig:CIRs}c) the bias tends to be much higher. In such case the receiver treats the first appearing component, which is usually one of reflected multipath components, as the direct path one. It introduces a large delay of several nanoseconds, which leads to a ranging error of even a few meters. This bias is very hard to mitigate, because it is almost impossible to determine the path, which the received component traveled. In such case the safest option would be to exclude those results from location calculation.

To properly mitigate or exclude deteriorated measurement results, NLOS conditions should be properly identified first. Most of the methods described in the literature distinguish only LOS and NLOS conditions. However due to higher biases of results obtained when direct path component is not present and difficulties with their mitigation it would be better to recognize two different situations: Direct-Path NLOS (DP-NLOS) when that component is available and Non-Direct-Path NLOS (NDP-NLOS) when it is completely blocked.

\section{Proposed approach} \label{approach}
The proposed approach to NLOS identification utilizes the features of Decawave's DW1000 IC \cite{decawave_ltd_dw1000_2015}, which is an IEEE 802.15.4a compliant radio chip. The employed IC has functionalities allowing to estimate the received signal power, power of first path component and to measure Channel Impulse Response, which in this case is a received signal. The signal recorded by the chip can be described as follows:
\begin{equation}
r(t) = \sum_{i=1}^{N}a_ip(t-\tau_i) + n(t)
\label{eq:CIR}
\end{equation}
where $N$ is the number of multipath components, $a_i$ and $\tau_i$ are the amplitude and delay of i\textsuperscript{th} component, $p$ is pulse waveform transmitted by the modules, and $n(t)$ is additive white Gaussian noise. The signal recorded by the chip can be used to extract various parameters which can be used in NLOS identification.

In the proposed method propagation conditions classification is performed based on the following signal features:
\begin{enumerate}[\IEEEsetlabelwidth{12)}]
\item Received Signal Power Level calculated as \cite{decawave_ltd_dw1000_2015} 
\begin{equation}
RSL = 10\log{10}{\left(\frac{C\cdot2^{17}}{N^2}\right)}-A  [dBm]\
\label{eq:RSL}
\end{equation}
where $C$ and $N$ are Channel Impulse Response Power value and Preamble Accumulation Count reported in  DW1000 registers and $A$ is a parameter dependent on pulse repetition frequency. 

\item Received Signal Power to First Path Power level ratio
\begin{equation}
RFPR = RSL-FSL\
\label{eq:RFR}
\end{equation}
where $FSL$ is estimated First Path Signal Power Level calculated as \cite{decawave_ltd_dw1000_2015}:
\begin{equation}
FSL = 10\log{10}{\left(\frac{F_1^2 + F_2^2+F_3^2}{N^2}\right)}-A  [dBm]\
\label{eq:FSL}
\end{equation}
where $F_1$, $F_2$ and $F_3$ are First Path Amplitude magnitude values reported in DW1000 registers.

\item signal energy
\begin{equation}
\epsilon_r = \int_{T}\left|r(t)\right|^2 dt\
\label{eq:energy}
\end{equation}
where $T$ is the time range in which the signal was registered.

\item mean excess delay spread
\begin{equation}
\tau_{MED} = \int_{T}t\frac{\left|r(t)\right|^2}{\epsilon_r} dt
\label{eq:me_ds}
\end{equation}

\item root mean square delay spread
\begin{equation}
\tau_{RMS} = \int_{T}(t-\tau_{MED})^2\frac{\left|r(t)\right|^2}{\epsilon_r} dt
\label{eq:rms_ds}
\end{equation}

\item mean value
\begin{equation}
\mu_{\left|r\right|} = \frac{1}{T}\int_{T}\left|r(t)\right| dt
\label{eq:mean}
\end{equation}

\item variance
\begin{equation}
\sigma_{\left|r\right|}^2 = \frac{1}{T}\int_{T}\left[\left|r(t)\right|-\mu_{\left|r\right|}\right]^2 dt\
\label{eq:var}
\end{equation}

\item kurtosis
\begin{equation}
\kappa = \frac{1}{\sigma_{\left|r\right|}^4T}\int_{T}\left[\left|r(t)\right|-\mu_{\left|r\right|}\right]^4 dt\
\label{eq:kurtosis}
\end{equation}

\item amplitude
\begin{equation}
A = max(\left|r(t)\right|)\
\label{eq:amp}
\end{equation}
\item variance of signal before First Path Component
\begin{equation}
\sigma_{s}^2 = \frac{1}{\tau_s}\int_{T_{FP}-\tau_s}^{T_{FP}}\left[\left|r(t)\right|-\mu_{\left|s(t)\right|}\right]^2 dt\
\label{eq:noise_var}
\end{equation}
where $\tau_s$ is length of a time frame in which the signal before the detection of First Path component ($T_{FP}$)  is analyzed and $\mu_{\left|s(t)\right|}$ is the average value of that signal.
\end{enumerate}

The first two parameters are easily attainable, because RSL and FSL are measured by DW1000 automatically and require only extraction from specific registers. That makes them perfect for real time NLOS identification methods.

Parameters 3--10 are calculated based on the recorded signals. Since their computation is more complex it's hard to apply them in real time. Parameters 3--9 are commonly used in many identification and mitigation methods . However the authors have not found the publication in which the variance of the signal directly preceding first path component identification was utilized. 

\begin{figure}[!h]
\centerline{\includegraphics[width=3.2 in]{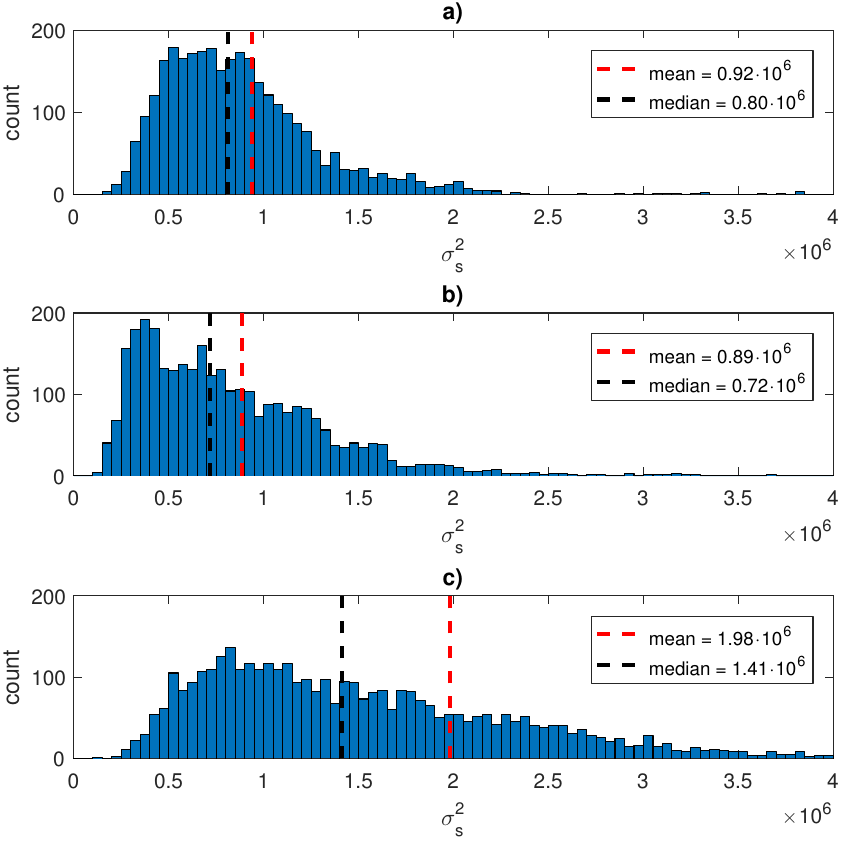}}
\caption{Histograms of variance of signal preceding first path detection in: a)~LOS  b) DP-NLOS, c) NDP-NLOS }
\label{fig:noiseHist}
\end{figure}

Analyzing variance of signal directly preceding first path component allows to identify situations as in Fig.\ref{fig:CIRs}.c, where heavily attenuated direct path component, which the chip is unable to properly detect is present. In such cases the variance is visibly higher. In order to capture that effect the signal should be analyzed over an adequate period of time. It was assumed that  $\tau_s$ value of 20 ns would be sufficient (the bias in those situations rarely exceeds that value). To evaluate working conditions influence on variance values, 10,000 waveforms were recorded under LOS, DP-NLOS and NDP-NLOS. The histograms of obtained variance values are presented in Fig.\ref{fig:noiseHist}.

The mean and median values of calculated variance for LOS (Fig.\ref{fig:noiseHist}a) and DP-NLOS (Fig.\ref{fig:noiseHist}b) are close so it would be very hard to recognize those conditions based on this parameter. However for NDP-NLOS the obtained histogram is different, the parameter achieves values from a larger range and mean and median are about two times higher than in case of the previous two. It seems that based on this parameter it would be possible to identify NDP-NLOS from DP-NLOS.

\section{NLOS identification} \label{ML}
\begin{figure}[!t]
\centerline{\includegraphics[width=2.3 in]{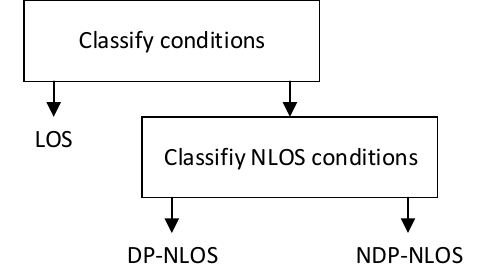}}
\caption{Classification method workflow }
\label{fig:class_me}
\end{figure}

Propagation conditions are classified  with Support Vector Machine (SVM) into one of three classes: LOS, DP-NLOS and NDP-NLOS based on the parameters listed in section \ref{approach}. In the paper a two-step classification method is proposed (Fig.\ref{fig:class_me}).

In the developed method working conditions are firstly identified to be LOS or NLOS, then for waveforms identified as NLOS another test is performed to check, whether the direct path component is available or not.

The SVM is a supervised   machine learning technique which is widely used to solve various classification problems \cite{marano_nlos_2010-1}. The basic version of SVM is a binary classifier, hence in the presented problem it is used in two-step classification.

Training phase of the algorithm consists in determining a hyperplane separating both categories. The resulting hyperplane is chosen in such a way that a margin between both categories is as wide as possible, which in case of limited training data allows to preserve algorithm's robustness. Additionally SVM is computationally inexpensive, which makes it a  suitable choice  to identify NLOS  in real time.

\section{Experiment results} \label{exp}

The proposed propagation conditions identification method was tested experimentally. The experiment consisted in collecting a large number of signal waveforms, RSL and FSL values in different propagation conditions and using them to verify the proposed method.

The measurements were carried out in a typical fully furnished apartment. To collect the waveforms Decawave's EVK1000 Evaluation kit was used \cite{decawave_ltd_evk1000_2015}. The kit consists of two evaluation boards EVB1000 containing DW1000 radio chips which are compliant to IEEE 802.15.4a standard. Both boards were connected to computers and were controlled using Decaranging application, which enabled to perform ranging measurements using Symmetric-Double-Sided Two-Way-Ranging (SDS-TWR) method \cite{albaidhani_ranging_2016-1}.

During the experiments the boards were placed in different locations in the apartment and  ranging procedure between them was performed. The measurements were taken in 57 different geometrical configurations. For each pair of points a couple of hundreds waveforms were collected which yielded a total of nearly 19,000 waveforms.

For each pair of points ranging bias was estimated and used to classify the propagation conditions into one of the three categories. Results with bias lower than 5 cm were assumed to be collected in LOS conditions. Based on the analysis of the apartment construction (concrete walls) and present furnishings it was presumed that the maximum bias introduced in DP-NLOS conditions would not exceed 70 cm. For higher biases  propagation conditions were classified as NDP-NLOS.

The collected data were divided into two groups containing samples obtained under all above propagation conditions. The first group comprised results for 24 pairs of points (ca. 7000 waveforms) and was used to train the algorithm. The rest of results (ca. 12000 waveforms) was used for testing.

The obtained measurement results were classified using SVM. The success rates (the ratio of correctly identified waveforms to total number) for LOS ($P_{LOS}$), NLOS ($P_{NLOS}$) and overall success rate defined as $P_o = (P_{LOS}+P_{NLOS})/2$ for different signal features are presented in Tab.\ref{tab1}.

The highest success rate of NLOS classification was achieved using $RSL$ and $RPFR$ (1,2). Unfortunately the number of properly classified LOS waveforms (0.581) is not satisfactory. The best results were obtained for parameters: $RFPR$, $\tau_{MED}$, $\tau_{RMS}$ (2,4,5). Waveforms classified as NLOS based on these features were used in the second step. 

The success rates for DP-NLOS/NDP-NLOS identification are stored in Tab.\ref{tab2}. Using the same data (2,4,5) to discern between NLOS scenarios does not yield in satisfactory results (overall success rate about 50\%). The highest NDP-NLOS identification and overall success rate were achieved based on signal energy and mean excess delay spread. However for that parameters a large number of DP-NLOS was not classified properly. Better DP-NLOS classification rates is possible based only on $\sigma_n$ (\ref{eq:noise_var}). In this situation using above parameters (3,4,10 combination) seems to be a good idea as it results in better overall consistency.

\begin{table}[t]
\renewcommand{\arraystretch}{1.3}
\caption{Success rates for LOS/NLOS classification}
\label{tab1}
\centering
\begin{tabular}{c||c|c|c}
\hline
\bfseries Used Features & \bfseries $P_{LOS}$ & \bfseries $P_{NLOS}$ & \bfseries $P_{avg}$ \\
\hline\hline
1,2 & 0.5818 & \textbf{0.9988} &0.7903\\
1,2,3 & 0.4337 & 0.9831 &0.7084\\
2,4,5 & \textbf{0.9472} & 0.9263 &\textbf{0.9367}\\
1,2,3,4,5 & 0.6508 & 0.9701 &0.8105\\
1,2,3,4,5,6,7,8,9 & 0.6522 & 0.9522 &0.8022\\
1,2,3,4,5,6,7,8,9,10 & 0.4799 & 0.9721 &0.7260\\
\hline
\hline
\end{tabular}
\end{table}

\begin{table}[t]
\renewcommand{\arraystretch}{1.3}
\caption{Success rates for DP/NDP classification}
\label{tab2}
\centering
\begin{tabular}{c||c|c|c}
\hline
\bfseries Used Features & \bfseries $P_{DP}$ & \bfseries $P_{NDP}$ & \bfseries $P_{avg}$ \\
\hline\hline
10 &  \textbf{0.8913} & 0.5549 & 0.7231\\
1,2 & 0.5124 & 0.5573 & 0.5349\\
2,4,5 & 0.6671 & 0.3976 & 0.5323\\
3,4 & 0.6120 & \textbf{0.9059} & \textbf{0.7590}\\
3,4,10 &   0.7322 & 0.6601 & 0.6961\\
1,2,3,4,5,6,7,8,9,10 & 0.6866 & 0.1929 & 0.4398\\
\hline
\hline
\end{tabular}
\end{table}

\section{Conclusion} \label{con}

In the paper an NLOS identification method has been proposed. The main novelty of the method is that it discerns a situation when the delayed direct path component is available from when it's totally blocked.

The method classifies propagation conditions using SVM algorithm based on various signal features. In addition to commonly used parameters, the method makes use of a variance calculated for signal directly before first path component detection.

The performed experiment has shown that the method allows for precise LOS/NLOS identification and identifies the absence of a direct path component with good accuracy.

The method can be possibly upgraded by employing another signal features and using more advanced machine learning techniques.

\bibliography{IEEEabrv,Mikon2018Bib}

\end{document}